\documentclass{article}
\usepackage[utf8]{inputenc}
\usepackage[letterpaper]{geometry}
\usepackage[parfill]{parskip}
\PassOptionsToPackage{numbers, compress}{natbib}
\usepackage{natbib}
\usepackage{xr-hyper,refcount}
\usepackage{graphicx}
\usepackage[utf8]{inputenc} % allow utf-8 input
\usepackage[T1]{fontenc}    % use 8-bit T1 fonts
\usepackage[colorlinks, citecolor=blue, urlcolor=blue, linkcolor=blue]{hyperref}       % hyperlinks
\usepackage{url}            % simple URL typesetting
\usepackage{booktabs}       % professional-quality tables
\usepackage{amsfonts}       % blackboard math symbols
\usepackage{nicefrac}       % compact symbols for 1/2, etc.
\usepackage{microtype}      % microtypography
\usepackage[parfill]{parskip}
\usepackage{algorithm,algorithmic}
\usepackage{amsmath,amsthm,amssymb,bbm}
\usepackage{mathtools}
\usepackage{cases}
\usepackage{comment}
\usepackage{subcaption}
\usepackage{algorithm,algorithmic}
\usepackage{color}
\usepackage{appendix}
\usepackage{xspace}
\usepackage{enumitem}
\usepackage[english]{babel}
\usepackage{authblk}

\usepackage{libertine}
\usepackage{libertinust1math}
\usepackage[T1]{fontenc}

\title{Algorithmic Fairness from a Non-ideal Perspective}

\author{Sina Fazelpour \& Zachary C. Lipton\\
Carnegie Mellon University \\
\href{mailto:sinaf@andrew.cmu.edu}{sinaf@andrew.cmu.edu}, \href{zlipton@cmu.edu}{zlipton@cmu.edu}   
}
\begin{document}

\maketitle

\begin{abstract}
Inspired by recent breakthroughs in predictive modeling,
practitioners in both industry and government 
have turned to machine learning with hopes 
of operationalizing 
predictions to drive automated decisions. 
Unfortunately, many social desiderata concerning
consequential decisions, such as justice or fairness,
have no natural formulation within a purely predictive framework.
In efforts to mitigate these problems,
researchers have proposed a variety of metrics
for quantifying deviations from various statistical parities
that we might expect to observe in a fair world
and offered a variety of algorithms
in attempts to satisfy subsets of these parities or
to trade off the degree to which they are satisfied against utility.
In this paper, we connect this approach to \emph{fair machine learning}
to the literature on ideal and non-ideal
methodological approaches in political philosophy.
The ideal approach requires positing the principles 
according to which a just world would operate.  
In the most straightforward application of ideal theory,
one supports a proposed policy by arguing that 
it closes a discrepancy between the real and the perfectly just world.
However, by failing to account for the mechanisms
by which our non-ideal world arose,
the responsibilities of various decision-makers,
and the impacts of proposed policies,
naive applications of ideal thinking 
can lead to misguided interventions.
In this paper, we demonstrate a connection
between the fair machine learning literature
and the ideal approach in political philosophy,
and argue that the increasingly apparent shortcomings 
of proposed fair machine learning algorithms
reflect broader troubles faced by the ideal approach.
We conclude with a critical discussion 
of the harms of misguided solutions, a reinterpretation of impossibility results, and directions for future research.\footnote{A version of this paper was accepted at the AAAI/ACM Conference on Artificial Intelligence, Ethics, and Society (AIES) 2020.}
\end{abstract}

% \keywords{algorithmic decision-making, 
% fairness in machine learning, 
% political philosophy, 
% justice
% }

%%%%%%%%%%%%%%%%%%%% Main
\section{Introduction}
\label{sec:introduction}
Machine Learning (ML) models play increasingly prominent roles
in the allocation of social benefits and burdens
in numerous sensitive domains, including hiring, social services, and criminal justice 
\citep{Berk2018,Crawford2014,Barocas2016,Feldman2015}.
A growing body of academic research and investigative journalism 
has focused attention on ethical concerns regarding
algorithmic decisions \citep{Brown2019,Dwork2012,Angwin2016}, 
with many scholars warning that in numerous applications, 
ML-based systems may harm members of 
already-vulnerable communities \citep{Barocas2016,Eubanks2018}. 

Motivated by this awareness, a new field of technical research 
addressing fairness in algorithmic decision-making has emerged, 
with researchers publishing countless papers aspiring to 
(i) formalize ``fairness metrics''---mathematical 
expressions intended to quantify the extent to which
a given algorithmic-based allocation is (un)just; 
and (2) mitigate ``unfairness'' as assessed by these metrics 
via modified data processing procedures, objective functions, 
or learning algorithms \citep{Hardt2016,Zafar2017,Nabi2018,Kilbertus2017,Dwork2012,Grgic-Hlaca2018a,Corbett-Davies2018}. 
However, progress has been hindered by disagreements 
over the appropriate conceptualization 
and formalization of fairness \citep{Chouldechova2016,Kleinberg2017,Glymour2019,Binns18}. 

The persistence of such disagreements raises 
a fundamental methodological question 
about the appropriate approach for constructing tools 
for assessing and mitigating potential injustices of ML-supported allocations. 
Importantly, any useful methodology must provide normative guidance
for how a given agent ought to act 
in a world plagued by systemic injustices.
Broadly speaking, justice requires apportioning benefits 
and burdens in accordance with each person's 
rights and deserts---giving individuals ``their due'' \citep{Miller2017,Feldman2016}.
Beyond this general framing, 
how can we offer more specific and practical guidance? 

Drawing on literature in political philosophy,
in Section \ref{sec:ideal-non-ideal},
we distinguish between \textit{ideal} and \textit{non-ideal} 
methodological approaches to developing such normative prescriptions,
and highlight three challenges facing the ideal approach. 
Then, in Section \ref{sec:fair-ml-ideal},
we argue that most of the current technical approaches 
for addressing algorithmic injustice 
are reasonably (and usefully) characterized 
as small-scale instances of ideal theorizing. 
% Following this argument, we relate the difficulties 
% faced by these attempts to operationalize fairness algorithmically 
% to the troubles that confront naive applications of 
% ideal theorizing more generally. 
Next, in Section \ref{sec:fair-ml-trouble},
we support this argument by demonstrating 
several ways that current approaches are, to varying extents, 
plagued by the same types of problems 
that confront naive applications of ideal theorizing more generally.
Finally, drawing on these considerations,
in Section \ref{sec:discussion},
we provide a critical discussion of 
the real-world dangers of this flawed framing
and offer a set of recommendations 
for future work on algorithmic fairness.

\section{Two Methodologies: Ideal \textit{vs.} Non-Ideal}
\label{sec:ideal-non-ideal}
How should one go about developing normative prescriptions 
that can guide decision-makers who aspire to act justly in an unjust world? 
A useful distinction in political philosophy is between
\textit{ideal} and \textit{non-ideal} modes of theorizing 
about the relevant normative prescriptions \citep{Simmons2010,Valentini2012,Stemplowska2012}.
When adopting the ideal approach, one starts
by articulating a conception of an ideally just world 
under a set of idealized conditions. 
The conception of the just world serves two functions: 
(i) it provides decision-makers with a \emph{target} state 
to aspire towards \citep{Stemplowska2012};
and (ii) when suitably specified, it serves 
as an \emph{evaluative standard} for identifying and assessing 
current injustices ``by the extent of the deviation 
from perfect justice'' \citep[p.~216]{Rawls1999}. 
According to this perspective, 
a suitably-specified evaluative standard 
can provide decision-makers with normative guidance 
to adopt policies that minimize deviations 
with respect to some notion of similarity,
thus closing the gap between the ideal and reality \citep{Anderson2010}.

Non-ideal theory emerged within political philosophy 
as a result of a number of challenges 
to ideal modes of theorizing \citep{Galston2010,Valentini2012}. 
We focus here on three challenges 
that motivate the non-ideal approach.
A first set of issues arises when we consider 
the intended role of a conception of an ideally just world 
as an evaluative lens for diagnosing actual injustices. 
In the ideal approach, the conceptual framing of perfect justice 
determines whether some actual procedure or event 
is identified \textit{as unjust} and if so, 
how that injustice gets represented \citep{Mills2005,Pateman2007}.
When this conception is impoverished,
e.g., by failing to articulate important factors,
it can lead to systematic neglect of injustices
that were overlooked in constructing the ideal. 
Moreover, the static nature of ideal standards 
and the pursuant diagnostic lens can overlook the factors 
that give rise to injustice in the first place.
This is because such standards identify injustices
in terms of \emph{the discrepancies} between the actual world 
and an ideally-just target state.
However, the historical origins and dynamics 
of current injustices and the ongoing social forces 
that sustain them are typically absent from consideration. 
By obfuscating these \emph{causal factors}, 
ideal evaluative standards can distort 
our understanding of current injustices.

According to a second challenge, employing a conception 
of an ideally just world as an evaluative standard 
is \emph{not sufficient} for deciding 
how actual injustices should be mitigated \citep{Sen2006,Sen2009}. 
This is because, from the standpoint of an ideal, 
any discrepancy between our imperfect world and that ideal
might be interpreted naively as a cause of an actual injustice, 
and thus, 
any policy that aims to directly minimize such a discrepancy 
might be erroneously argued to be justice-promoting \citep{Anderson2010,Sen2006}.
Yet, the actual world can deviate from an ideal in multiple respects,
and the same kind of deviation can have 
varied and complex causal origins \citep{Sen2009}.
Moreover, as the fair machine learning literature clearly demonstrates (see Section \ref{sec:impossibility}),
simultaneously eliminating all discrepancies might be impossible. 
Thus, a coherent approach requires 
not only a mandate to eliminate discrepancies, 
but also guidance for determining which discrepancies matter in a given context. 
Crucially, policies that simply seek to minimize 
any perceived gap between the ideal and reality without 
consideration for the underlying causes
may not only be ineffective solutions to current injustices, 
but can potentially exacerbate the problem they purport to address.
For example, ideal theorizing has been applied 
to argue for race-blind policies (against affirmative action) \citep{Anderson2010}.
From the perspective of an ideally just society as a race-blind one, 
a solution to current injustices 
``would appear to be to end race-conscious policies'' \citep[4]{Anderson2010}, 
thus blocking efforts devised to address historical racial injustices.
Absent considerations of the dynamics by which disparities emerge,
it is not clear that in a world where individuals have been racialized
and treated differently on account of these perceived categories,
race-blind policies are capable of bringing about the ideal \citep{Anderson2010}.

Finally, a third challenge concerns the practical usefulness 
of the ideal approach \textit{for current decision-makers},
given the type of idealized assumptions 
under which ideal theorizing proceeds.
Consider, for example, the assumption of \textit{strict compliance},
frequently assumed by ideal theorists as a condition
under which the conception of an ideally just world can be developed.
The condition assumes that nearly all relevant agents
comply with what justice demands of them \citep[13]{Rawls2001}.
The condition thus idealizes away situations
where some agents fail to act in conformity with their ethical duties 
(e.g., the duty not to racially discriminate), or are unwilling to do so. 
The vision of a just world constructed under this assumption 
fails to answer questions about what we might reasonably expect 
from a decision-maker in the real world, 
where others often neglect or avoid their responsibilities 
\citep{Schapiro2003,Feinberg1973,Valentini2012}.

In short, when used as lens for identifying current injustices, 
ideal modes of theorizing 
(1) can lead to systematic neglects of some injustices
and distort our understanding of other injustices; 
(2) do not, by themselves, offer sufficient practical guidance about \emph{what should be done}, 
sometimes leading to misguided mitigation strategies;
and finally, (3) do not, by themselves, 
make clear \emph{who, among decision-makers} 
is responsible for intervening to right specific injustices.
As a result of these challenges to ideal modes of theorizing, 
a number of researchers in political philosophy 
have turned to non-ideal modes of theorizing. 
In contrast to the ideal approach, 
the non-ideal approach begins by identifying 
actual injustices that are of concern to decision-makers 
and that give rise to reasonable complaints 
on behalf of those affected by their decisions \citep{Anderson2010,Sen2006}. 
Non-ideal theorizing can be seen as a trouble-shooting effort 
towards addressing these actual concerns and complaints. 
As Sen notes, this trouble-shooting aim distinguishes 
non-ideal modes of theorizing from ideal approaches 
that focus ``on looking only for the simultaneous fulfilment 
of the entire cluster of perfectly just societal arrangements'' \citep[p.~218]{Sen2006}. 

Anderson offers a succinct description of the non-ideal
approach towards this trouble-shooting goal
and what that approach requires:
\begin{quote}
    [Non-ideal theorists] ... seek a causal explanation of the problem 
    to determine what can and ought to be done about it,
    and who should be charged with correcting it.
    This requires an evaluation of the mechanisms causing the problem,
    as well as responsibilities of different agents to alter these mechanisms \citep[p.~22]{Anderson2010}
\end{quote}
As noted by Anderson, there is still a crucial role
for normative ideals within the non-ideal approach.
But this role is importantly different
from the roles assigned to ideals 
in the ideal approach \citep[6]{Anderson2010}.
In the ideal approach, normative ideals are \textit{extra-empirical},
in the sense that they set the evaluative standards
against which actual practices are assessed,
without themselves being subject to empirical evaluation.
In contrast, in non-ideal theorizing, 
normative ideals act as \textit{hypotheses}
about potential solutions to identified problems. 
Viewed in this way, normative ideals are subject to revision 
in light of their efficacy in addressing
the concerns and complaints that arise in practice.
In the following sections, 
we show how the distinction can be put to work 
in understanding and addressing algorithmic injustice.

\section{Work on Algorithmic Fairness as Small-scale Ideal Theorizing}
\label{sec:fair-ml-ideal}
In political philosophy, the distinction 
between ideal and non-ideal approaches 
typically refers to ways of understanding
the demands of justice at large, 
and offering practical normative guidance
to basic societal institutions 
for complying with these demands. 
While some researchers are beginning to discuss 
how the automation of decision making in consequential domains 
interacts with demands of justice at this large scale, 
most works on algorithmic fairness have the more restricted aim
of assessing and managing various disparities that arise 
among particular demographic groups in connection 
with the deployment of ML-supported decision systems
in various (often-allocative) settings.
Nonetheless, in what follows, we show that 
the distinction between ideal and non-ideal approaches 
provides a fruitful lens for formulating strategies 
for addressing algorithmic injustices, 
even on this smaller scale (of an individual decision-maker). 
In this section, we argue that 
the dominant approach among current efforts towards addressing algorithmic harms 
can be seen as exercises in \emph{small-scale} ideal theorizing. 

\subsection{Developing a Fairness Ideal}
Works on algorithmic fairness typically begin 
by outlining a conception of a ``fairness ideal''. 
\citet[p.~215]{Dwork2012}, for example, seek to ``capture fairness 
by the principle that any two individuals 
who are similar with respect to a particular task should be classified similarly'' (see also \citet{Jung2019}). 
Others envision the fair ideal at the group level. 
In nearly all cases, the groups of interest 
are those encompassing categories such as race, 
ethnic origin, sex, and religion. 
Following precedent in the United States Civil Rights Act, 
these groups are typically called \emph{protected classes} 
or \emph{protected groups} in the technical literature. 
According to one group-level conception of fairness, 
fair allocative policies and procedure are those 
that result in outcomes that impact different protected groups 
in the same way \citep{Zafar2017,Feldman2015}. 
In other cases, a fair state is taken to be one 
in which membership in a protected group is irrelevant 
or does not make a difference to the allocative procedure \citep{Kilbertus2017,Grgic-Hlaca2018a}. According to another view, a treatment disparity might exist in a fair state, 
if it is justified by the legitimate aims of the distributive procedure \citep{Hardt2016,Nabi2018}. 
The endorsed fairness ideals have different provenances:
in some cases, authors refer to historical legal cases, 
such as \textit{Carson v. Bethlehem Steel Corp.} or \textit{Griggs v. Duke Power}, 
to support their conception of fairness. 
In other cases, the ideal of fairness is derived from people's intuitive judgments 
about fair allocation \citep{Grgic-Hlaca2018a,Jung2019}. 
And less frequently, authors allude to works of political philosophers 
such as Rawls, which is cited to support 
the conception of individual fairness in \citet{Dwork2012}.

\subsection{Specifying a Fairness Metric}
Next, on the basis of their favored fairness ideal, 
researchers specify a quantitative evaluative standard---a ``fairness metric''---for 
diagnosing potential allocative injustices and guiding mitigation efforts. 
Typically, these fairness metrics take the form of mathematical expressions 
that quantify how far two among the protected groups are from \emph{parity}. 
The magnitude of (dis)parity measured by a given fairness metric 
is taken to denote the degree of divergence from the ideal
for which that metric is supposed to be a formal proxy. 

Given their generality and abstract nature, 
fairness ideals do not fully determine the specific shape of fairness metrics. 
Accordingly, in addition to a fairness ideal, the construction of fairness metrics 
requires researchers to make further value judgments.
For example, the ideal that membership in protected groups 
should be irrelevant to allocative decisions can be articulated 
in the language of statistics by requiring the outcome $\hat{Y}$  
be independent (probabilistically) of the protected attributes $A$ \citep{Feldman2015}. 
However, the same ideal can also be expressed in the language of causality, 
e.g., by requiring that the average causal effect 
of protected attributes $A$ on $\hat{Y}$ be negligible \citep{Kilbertus2017}.
Similarly, one can formalize the qualification 
that protected attributes can make a difference to outcomes 
when justified by the legitimate aims of allocative procedures in different ways. 
In the language of statistics, for example, one can require 
that while there may be some correlation between $\hat{Y}$ and $A$,
the dependency must be screened off by the target variable, $Y$ \citep{Hardt2016}.
Framed in the language of causality, 
some attempt to formalize this fairness ideal 
in terms of a parity between the causal effect of $A$ on $\hat{Y}$ 
along so-called \emph{legitimate pathways} \citep{Nabi2018}, 
where what counts as legitimate
depends on the specific task and $Y$. 
Importantly, despite being motivated by the same ideal, 
such fairness metrics make different demands from the user 
and can result in different verdicts about the same case. 
In general, while statistical metrics can be formulated 
as functions of the joint distribution $P(Y, \hat{Y}, A, X)$, 
causal metrics additionally require the acquisition of a causal model 
that faithfully describes the data-generating processes
and for which the desired causal effect is identifiable.
Thus in some situations, statistical parity metrics may be estimable from data
while the corresponding causal quantities may not be,
owing to our limited knowledge of the data-generating process \citep{pearl2009causality}.

\subsection{Promoting Justice by Minimizing Deviations from the Ideal}
Finally, current approaches seek to promote fairness (or mitigate unfairness) 
by modifying ML algorithms to maximize utility subject 
to a parity constraint expressed in terms of the proposed fairness metric. 
Such fairness-enforcing modifications can take the form of interventions 
(i) in the pre-processing stage to produce ``fair representations'' 
(e.g., \citet{Kamiran2012});
(ii) in the learning stage to create ``fair learning'' 
(e.g., \citet{Zafar2017}); or 
(iii) in the post-processing by adjusting the decision thresholds
(e.g., \citet{Hardt2016}).
Crucially, however, in all cases, the range of solutions to algorithmic harms 
is limited to an intervention \emph{to the ML algorithm}. 
Absent from consideration in these approaches is the broader context in which 
the ``certifiably fair'' model will be deployed.
Recalling Anderson's critique \citep*[22]{Anderson2010} of ideal approaches, 
neither the mechanisms causing the problem,
nor the consequences of algorithmically-guided decisions,
nor ``the responsibilities of different agents to alter these mechanisms''
are captured in any of these approaches.

\section{Troubles with Ideal Fairness Metrics}
\label{sec:fair-ml-trouble}
If current works on algorithmic fairness pursue (small-scale) ideal theorizing, 
then we should expect these works to encounter the same types of challenges 
as those confronting ideal theorizing more generally. 
As explained above, 
according to critics, ideal modes of theorizing can 
(1) lead to systematic neglects of some injustices; 
and distort our understanding of other injustices. 
Such ideal evaluative standards 
(2) do not offer sufficient practical guidance 
and can lead to misguided mitigation strategies. 
What is more, they (3) fail to delineate the responsibilities 
of current decision-makers in a world 
where others fail to comply with their responsibilities. 
Below, we consider each of these challenges in turn, 
and show that these same types of worries arise 
with respect to current works on algorithmic fairness. 

\subsection{Systematic Neglects of Rights}
The identification of injustices in ideal theorizing is constrained 
by the underlying conceptual framing of normative ideals. 
If this conceptual framing is not sufficiently rich or comprehensive, 
we run the risk of overlooking many actual injustices. 
The ideals of fairness in literature on algorithmic fairness 
are predominantly expressed in terms of 
some type of parity among designated protected classes.
Is this comprehensive enough to be sensitive to the types of injustices 
that would lead to legitimate complaints 
by those affected by ML-based allocations? 
We believe that the answer is negative. 
To see why, consider that 
assessing claims of injustice 
can require attending to different types of information. 
As noted by Feinberg \citep{Feinberg1974,Feinberg2014}, 
in some cases, what is someone's due 
is determinable only in comparison 
to what is allocated to others 
or what would have been allocated to them had they been present. 
In other cases, an individual's just due is determinable
independent of any comparison and solely by reference 
to how that individual should have been treated 
in light of her rights and deserts. 
An allocative procedure can thus result in \emph{comparative} 
as well as \emph{non-comparative} cases 
of injustice \citep{Feinberg1974,Feldman2016,Montague1980}.

Yet, virtually all algorithmic fairness ideals 
are framed in \textit{comparative terms}. 
This comparative focus renders these ideals insensitive 
to legitimate claims of non-comparative injustice. 
Consider from this perspective, a judge who treats all defendants equally, 
denying parole to them all regardless of the specifics of their cases. 
Here the defendants can feel aggrieved because of 
how they \emph{should have been} treated from the perspective 
of the standards of retributive justice; 
the review process was based on legally irrelevant factors, 
infringing on defendants’ rights to due process, 
and at least in some cases, 
the punishments were disproportionately harsh, 
potentially resulting in arbitrary incarceration. 
Indeed, such sentencing behaviour goes against Articles 9 and 11 
of the Universal Declaration of Human Rights, 
cited throughout various documents concerning ethical design 
such as the \emph{IEEE Ethically Aligned Design} 
and the Toronto Declaration \citep{IEEE}. 
Yet, this and other cases of non-comparative injustice 
in which an individual's rights and deserts have been ignored 
escape the purview of current fairness metrics.

The situation is troubling even with respect to 
\textit{comparative} cases of injustice. 
This is because, due to their narrow focus, 
fairness metrics essentially take the set of protected classes 
to \textit{exhaust} comparison classes that might matter 
from the perspective of justice and fairness. 
However, consider a case where the appraisal of an employee's performance 
is influenced by factors such as their weight or height, 
despite the irrelevance (in a causal sense) of such characteristics 
to that job \citep{Judge2004,Rudolph2009}.
In this setting and from the perspective of comparative justice, 
height and weight \textit{are} relevant categories. 
The complete reliance of such metrics on the particular specification 
of relevant comparison groups  
limits their adequacy in this regard. 
Indeed, unconstrained by these demands of comparative justice, 
algorithmic-based decisions might result 
in the creation of new ``protected groups''.

\subsection{Distortion of the Harms of Discrimination}
From the perspective of current fairness ideals, 
any divergence from the ideal of parity among protected classes 
(potentially subject to certain qualifications) 
is identified as a case of unfairness. 
Accordingly, the fairness metrics based on these ideals 
often have the property of being \emph{anonymous} or \emph{symmetric}; 
whether a distribution of benefits and burdens is fair 
does not depend on who the affected individuals or groups are.
In certain contexts and for certain purposes, anonymity is a desirable property. 
Quantitative metrics of \emph{income inequality} are required to be anonymous, 
for example, because ``from an ethical point of view, 
it does not matter who is earning the income'' \citep{Ray1998}. 
Unlike the case of income inequality, however, 
evaluating fairness claims requires going beyond 
the observation \emph{that} some disparity exists \citep{Hellman2008}. 
We need to know \emph{why} the disparity exists 
and to understand ``the processes that produce or maintain it'' \citep[18]{Anderson2010}. 
This knowledge is required to determine a coherent course of action,
and yet it does not inform any of the mitigation strategies 
in the standard fair machine learning tool-kits,
making them unsuitable for off-the-shelf application.

Consider, for example, the very different mechanisms
giving rise to disparities in representation between 
(white and east Asian) vs (white and black) students in US higher education. 
In the former case, the disparity 
(appearing to favor Asian students)
emerges despite historical and institutional discrimination.
In the latter, the disparity stems from well-documented 
historical and institutional discrimination.
However, both represent violations of demographic parity \citep{Petersen1976}.
A naive ideal approach may suggest that in both cases,
the disparity requires alterations in admissions policies 
to enforce the parity across all groups we might expect in our ideal.
A more nuanced non-ideal approach might recognize 
the differences between these two situations.
In the literature on fair ML,
approaches that incorporate knowledge of demographic labels
are colloquially referred to as ``fairness through awareness''.
However, as demonstrated above, 
awareness of demographic membership alone 
is too shallow to distinguish between these two situations.
Instead, we require a deeper awareness, 
not only of demographic membership 
but of the societal mechanisms that imbue demographic membership
with social significance in the given context
and that give rise to existing disparities. 

While this is especially problematic for statistical metrics 
that neglect the provenance of the observed data, 
recently-proposed causal approaches, 
including those formalizing fairness in terms of average causal effect 
or the effect of treatment on the treated, 
are similarly insufficient for capturing 
when a given disparity is reflective of discrimination,
let alone whose discrimination it might reflect
or providing guidance as to when the current decision-maker 
has a responsibility or license to intervene.
Importantly, these causal methods typically address 
the problem of mediation analysis, 
adopting the perspective of an auditor seeking
to explain the mechanisms by which the protected trait 
influences a model's prediction. 
Missing however, is a coherent theory 
for how to relate those mechanisms to the 
responsibilities of the current decision-maker,
or any accounting of the causal mechanisms by which
a proposed intervention may impact the social system
for better or worse.

\subsection{Insufficient Insights and Misguided Mitigation}
As noted in the previous section, current mitigation strategies 
are guided by the idea that justice is promoted 
by intervening on ML algorithms to minimize 
disparities detected by a given metric. 
Insofar as the underlying causes of preexisting disparities 
and the consequences of proposed policies are ignored, however, 
these mitigation techniques might have adverse effects.
As one example, consider a series of proposed approaches 
that \citet{Lipton2018} denote \emph{disparate learning processes} (DLPs).
These techniques are designed to jointly satisfy two parities,
blindness and demographic parity (e.g., \citet{Zafar2017}).
However, as \citet{Lipton2018} (2018) show, 
DLPs are oblivious to the underlying causal mechanisms of potential disparities 
and in some cases, DLPs achieve parity between protected classes (e.g., genders) 
by giving weight to the irrelevant proxies, (e.g., hair length).
Using real-world data from graduate admissions to a computer science program,
they showed that prohibited from considering gender directly,
a DLP would pick up on proxies such as the subfield of interest.
In order to achieve parity, the DLP must advantage those applicants
that appear (based on their non-protected attributes)
to be more likely to be women, while disadvantaging 
those that are more likely to be men.
Thus, the DLP satisfies demographic parity 
by advantaging those pursuing studies
in sub-fields chosen historically by more women 
(e.g., human-computer interaction) 
while disadvantaging those pursuing studies
that are currently more male-dominated (e.g., machine learning).
While the DLP achieves overall demographic parity, 
women in fields that already have greater parity receive the benefit,
while women in those precise fields 
that most want for diversity
would actually be penalized by the DLP.

Stepping back from a myopic view of the statistical problem
and these arbitrarily-chosen deviations (the fairness metrics) from an ideal,
when we consider the impact of a deployed DLP on a broader system of incentives,
it becomes clear that the DLP risks amplifying
the very injustices it is intended to address.

In addition to the non-comparative harm of making decisions on irrelevant grounds, 
the supposed remedy can reinforce social stereotypes,
e.g., by incentivizing female applicants 
towards only those fields where 
they are already well represented (and away from others).
Similarly, in simply seeking to minimize 
the disparity detected by fairness metrics, 
current metrics neglect considerations 
about whether the enforced parity 
might in fact result in long term harms \citep{liu2018delayed}.

\subsection{Lack of Practical Guidance}
Finally, consider that the type of unjust disparities 
often faced in a given allocation context 
correspond to events potentially unfolding over decades. 
Current approaches to algorithmic fairness 
seek to address \emph{``is there discrimination?''} 
but leave open the questions of \emph{``who discriminated?''} 
and \emph{``what are the responsibilities of the current decision-maker?''} 
If sensitive features influence education, 
which in turn influences employment decisions,
then to what extent does the causal effect reflect the discrimination 
of the education system compared to that of the employer? 
The answer to this question is not straightforward
and requires considerations not captured in the entries of confusion matrices. 
While identifying statistical disparities may be valuable unto itself, 
e.g., as a first step to indicate particular situations that warrant investigation,
it provides little moral or legal guidance to the decision-maker.
While the influence of protected attributes on predictions may reflect injustice, 
providing normative guidance requires
identifying not only what would constitute a just world 
but also what constitute just decisions in the actual world,
with its history of injustice. 

%\section{Related Work}
%\label{sec:related}
%\input{sections/related.tex}

\section{Discussion}
% for Future Research}
\label{sec:discussion}
\subsection{If not Solutions, then Solutionism?}
Even as the mitigation strategies arising
from the recent technical literature on fair machine learning 
fail to offer practical guidance on matters of justice, 
they have not failed to deliver in the marketplace. 
From the perspective of stakeholders caught in the tension
between (i) the potential profit to be gained from deploying 
machine learning in socially-consequential domains,
and (ii) the increased scrutiny of a public 
concerned with algorithmic harms,
these metrics offer an alluring solution:
continue to deploy machine learning systems per the status quo,
but use some chosen parity metric to claim 
a certificate of fairness, seemingly inoculating the actor
against claims that they have not taken the moral concerns seriously,
and weaponizing the half-baked tools produced 
by academics in the early stages of formalizing fairness
as a shield against criticism.

In socially-consequential settings, requiring caution 
or even abstention (from applying ML)
such as criminal justice and hiring,
fair ML offers an apparent academic stamp of approval.
Notable recent examples include the IBM fairness 360 toolkit,
which offers fairness metrics and corresponding mitigation strategies 
as an open-source software service that claims to be able 
to ``examine, report, and mitigate discrimination and bias 
in machine learning models throughout the AI application lifecycle'' \citep{ibm360fairness}.
Using just one parity metric (demographic parity),
algorithmic hiring platform Pymetrics, Inc. 
claims that their system is ``proven 
to be free of gender and ethnic bias'' \citep{pymetrics}.

The literature on fair machine learning bears 
some responsibility for this state of affairs. 
In many papers, these fairness-inspired parity metrics 
are described as \emph{definitions of fairness}
and the resulting algorithms that satisfy the parities 
are claimed axiomatically to be \emph{fair}.
While many of these \emph{metrics} are useful diagnostics,
potentially alerting practitioners to disparities warranting further investigation,
the looseness with definitions creates an opening for stakeholders 
to claim compliance, even when the problems have not been addressed.
Lacking the basic primitives required to make the relevant moral distinctions,
when blindly optimized, these metrics 
are as likely to cause harm as to mitigate it. 
Thus current methods produced by the fair ML community
run the risk of serving as \emph{solutionism} 
if not as solutions \citep{Selbst2019}.

\subsection{Re-interpreting Impossibility Results}
\label{sec:impossibility}
An additional benefit of viewing fairness in ML
through the lens of non-ideal theorizing in political philosophy
is that it gives a new perspective for parsing 
the numerous impossibility results \citep{Kleinberg2017, Chouldechova2016}
famously showing that many statistical 
fairness metrics are irreconcilable, 
presenting inescapable trade-offs. 
These results are sometimes misinterpreted 
as communicating that \emph{fairness is impossible}.
However, through the non-ideal lens, 
these impossibility theorems
are simply a frank confirmation of the fact 
that we do not live in an ideal world. 
The inputs to statistical fairness metrics
include four groups of variables: 
the covariates $X$, the group membership $A$,
the label $Y$, and the classification $\hat{Y}$.
The distribution over these variables
at a given point in time 
is the consequence of the complex dynamics of an unjust society 
constituted of many decision-making agents. 
Of these, the current decision-maker has control 
only over their own predictions $\hat{Y}$.
That various metrics/parities cannot be satisfied simultaneously
merely by setting the values taken by $\hat{Y}$
indicates only that our present decision-maker cannot
\emph{through their actions alone} bring about 
the immediate end to all disparity, 
even as viewed locally through the variables
that their individual decisions concern.

One potential contribution of ML impossibility theorems
to philosophy is that they make evident 
an often-overlooked shortcoming with the ideal approach.
These impossibility results make clear 
that in general, if we start from a non-ideal world, 
no set of actions (by a single agent) 
can instantaneously achieve the ideal world in every respect.
Moreover, matching the ideal in a particular respect 
may only be possible at the expense of widening gaps in others. 
Thus this naive form of an ideal approach 
appears to be fundamentally under-specified.
If matching the ideal in various respects simultaneously is impossible, 
then we require, in addition to an ideal, a basis for deciding
which among competing discrepancies to focus on.
In this manner, the impossibility results in fair ML 
provide a novel lens to approach the philosophical debate 
about the extent to which normative theorizing on matters of justice 
can proceed in isolation from empirical socio-historical facts \citep{Sen2009,Farrelly2007}. 

While characterizing disparities and understanding 
the fundamental trade-offs among them may be valuable work, 
this work cannot by itself tell us what to do.
The pressing issue in determining how to act justly
is not how to optimize a given metric
but how to make the determination of what,
in a given situation, should be optimized in the first place.

\subsection{Towards a Non-Ideal Perspective}

Even if the reader finds the case against the ideal approach compelling,
there remains a pragmatic question of what precisely a non-ideal
approach might look like in practice. 
To begin, non-ideal theorizing about the demands of justice 
is a \emph{fact-sensitive} exercise.
Offering normative prescriptions to guide actions
requires understanding the relevant causal mechanisms 
that (i) account for present injustices; 
and (ii) govern the impact of proposed interventions.

\subsubsection*{Empirical understanding of the problem: }
Developing causal models for understanding
social dynamics that cause and maintain particular injustices 
requires extensive domain-knowledge 
as well as numerous value judgements 
about the relevance and significance of 
different aspects of the domain of interest.
Choices must be made about what abstractions are reasonable,
which simplifying assumptions are justified, 
and what formalizations are appropriate.
Inevitably, these choices, embedded in design and modeling,
raise \emph{coupled ethical-epistemic} questions \citep{Tuana2010,Proctor2008}.
Consider, for instance, choices that might be made 
in understanding the causes of racial injustice 
in a particular allocative domain and a specific social setting.
Aside from the challenge of understanding 
the concept of race \citep{Mills1998,Mallon2006}, 
research in psychology and sociology shows racial classification 
and identification to be dynamic categories 
that are shaped by a variety of socioeconomic factors 
such as unemployment, incarceration, and poverty 
\citep{Epp2014,Penner2008,Freeman2011}.
Appreciating the complex and dynamic nature of race and the perception thereof 
is thus not only of ethical import; it also has important epistemic implications 
for formal models of racial injustice, 
as it shapes how ``race'' as an attribute should be understood 
and what causal relation it might bear to other factors of interest.

\subsubsection*{Empirically-informed choice of treatment: }
Deployment of predictive models---whether 
those that simply maximize utility or those that maximize utility 
subject to some ``fairness'' constraint---constitutes a social intervention. 
As mentioned above, most existing approaches to fair ML 
consist only of modifying the data processing procedures or the objective functions. 
Crucially, the evaluation of these interventions is \emph{local} and \emph{static}: 
the evaluation is local insofar as it concerns the impact 
of the intervention only on that particular predictive model's statistics
(i.e., its accuracy and various fairness metrics).
The accompanying literature seldom  
considers the broader impacts of deploying such models 
in any particular social context. 
Moreover, the evaluation is typically static, 
ignoring the longer-term dynamics of proposed policies.
When authors have attempted dynamic evaluations,
the results have sometimes contraindicated proposed mitigation strategies
\citep{liu2018delayed}.

In contrast, a non-ideal approach to offering normative guidance 
should be based on evaluating the situated and system-wide 
(involving not just the predictive model 
but also the broader social context, actors, and users) 
and dynamic (evolving over longer periods) impact 
of potential fairness-promoting interventions.

% \textcolor{blue}{\cite{Selbst2019,Hoffmann2019} also emphasize the importance of evaluating the broader impact of ML systems in their deployment context}

Once more, we must face difficult questions and make value judgments. 
As some authors have noted, for instance,
unjust circumstances can naturally arise as a result 
of seemingly benign initial conditions \cite{Schelling1971,OConnor2019a}.
To determine how to act, a coherent framework is needed
for understanding when is it desirable or permissible 
for a given decision-maker to intervene. 
Importantly, we stress that 
the appropriate judgments 
simply cannot be made based on the reductive 
($X$, $A$, $Y$ $\hat{Y}$) description
upon which most statistical fair ML operates. 
Developing a coherent non-ideal approach
requires (for the foreseeable future) human thought, 
both to understand the social context 
and to make the relevant normative judgments.

\section{Conclusion}
\label{sec:conclusion}
Approaching the issue of algorithmic fairness from a non-ideal perspective requires a broadening of scope beyond parity-constrained predictive models, and considering the wider socio-technological system consisting of human users, who informed by these models, make decisions in particular contexts and towards particular aims. 
Effectively addressing algorithmic harms demands nothing short of this broader, human-centered perspective, as it enables the formulation of novel and potentially more effective mitigation strategies that are not restricted to simple modifications of existing ML algorithms.

\section*{Acknowledgements}
\label{sec:ackknowledge}
Many thanks to David Danks, Maria De-Arteaga, and our reviewers for helpful discussions and comments. Funding was provided by Social Sciences and Humanities Research Council of Canada (No. 756-2019-0289) and the AI Ethics and Governance Fund.

\bibliographystyle{plainnat}
\bibliography{refs}
\end{document}